\documentclass[12pt]{article}
\DeclareUnicodeCharacter{03A6}
{\ensuremath{\Phi}}

\usepackage{amsmath}

\usepackage{amssymb}

\usepackage{doi}

\usepackage{longtable}

\usepackage{graphicx}

\usepackage{float} 

\usepackage{cite}
\usepackage{tabularx}

\usepackage{microtype} 

\usepackage[utf8]{inputenc}
\usepackage{longtable}
\usepackage{array}
\usepackage{ragged2e}
\usepackage[utf8]{inputenc}
\usepackage{enumitem}
\usepackage{xurl}

\usepackage{float}

\title{Proof of Humanity: A Multi-Layer Network Framework for Certifying Human-Originated Content in an AI-Dominated Internet}
\author{Sebastian Barros}
\date{April 1st , 2025}  

\begin{document}
\renewcommand{\arraystretch}{1.2}

\maketitle

\begin{abstract}
The rapid proliferation of generative AI has led to an internet increasingly populated with synthetic content—text, images, audio, and video generated without human intervention. As the distinction between human and AI-generated data blurs, the ability to verify content origin becomes critical for applications ranging from social media and journalism to legal and financial systems.

In this paper, we propose a conceptual, multi-layer architectural framework that enables telecommunications networks to act as "infrastructure level certifiers" of human-originated content. By leveraging identity anchoring at the physical layer, metadata propagation at the network and transport layers, and cryptographic attestations at the session and application layers, Telcos can provide an end-to-end "Proof of Humanity" for data traversing their networks. We outline how each OSI layer can contribute to this trust fabric using technical primitives such as SIM/eSIM identity, digital signatures, behavior-based ML heuristics, and edge-validated APIs.

The framework is presented as a foundation for future implementation, highlighting monetization pathways for Telcos such as trust-as-a-service APIs, origin-certified traffic tiers, and regulatory compliance tools. The paper does not present implementation or benchmarking results, but offers a technically detailed roadmap and strategic rationale for transforming Telcos into validators of digital authenticity in an AI-dominated internet. Security, privacy, and adversarial considerations are discussed as directions for future work.

\end{abstract}

\section{Introduction}

\subsection{Context and Urgency}

The exponential growth of generative artificial intelligence (AI) is transforming the nature of digital content. Large Language Models (LLMs) such as GPT-4 \cite{openai2023gpt4}, image generators like Midjourney and Stable Diffusion \cite{rombach2022high}, and emerging multimodal systems (e.g., Sora by OpenAI) now produce text, images, video, and audio at scale, often indistinguishable from human-created media. 

Recent projections suggest that AI-generated content could account for up to 90\% of all internet data by the end of this decade. This inflection point introduces a fundamental challenge: the loss of trust in the authenticity and origin of digital content. Malicious actors can now create synthetic identities, deepfakes, and automated propaganda at near-zero cost. Even benign AI usage can saturate platforms with generated content, making it difficult to distinguish what is human and what is machine.

This surge in synthetic data erodes public trust, weakens the integrity of social platforms and search engines, and creates risks for legal, financial, and governmental decision-making systems that rely on presumed human intent or authorship.

\subsection{Problem Statement}

Despite the rapid advancement of generative AI, the global digital infrastructure lacks a scalable, neutral mechanism to determine whether content was created by a human or an autonomous system. Most current approaches focus on the application layer. For instance, cryptographic watermarking, metadata signing (e.g., C2PA), and AI-generated disclaimers rely heavily on voluntary participation by content creators or platforms~\cite{c2pa2023spec}. These solutions are fragile—metadata can be stripped, watermarks can be removed, and platform-specific implementations create fragmentation across the internet.

On the other end of the spectrum, detection-based methods attempt to identify AI-generated content using classifiers or statistical signatures~\cite{tiwari2024detecting}. However, these are inherently reactive and adversarial: generative models evolve rapidly, often outpacing detection capabilities. Moreover, detection algorithms tend to produce false positives and false negatives, especially on short or ambiguous inputs, and are unreliable in high-stakes domains like journalism, legal evidence, or political communication.

What is missing is a robust, infrastructure-level method to propagate and verify content origin. A mechanism not tied to a single app or cloud provider—but embedded in the underlying network fabric that delivers content. Such a mechanism must be scalable, protocol-agnostic, privacy-aware, and resistant to manipulation. Critically, it must allow for the verification of human origin in real time, without requiring payload inspection.

This paper argues that telecommunications networks, by virtue of their trusted subscriber relationships and full-path data control, are in a unique position to provide such infrastructure. We propose a framework in which Telcos enable multi-layer “Proof of Humanity” by embedding and verifying identity credentials at multiple points in the OSI stack, thereby restoring trust in the origin of digital content as it transits through the network.

\subsection{Opportunity for Telcos}

Telecommunication networks operate at the foundational layers of the internet stack. Unlike application providers or AI platforms, Telcos maintain persistent, regulated, and identity-bound relationships with end-users through SIM provisioning, device registration, and subscription-based billing infrastructures. These relationships are often backed by Know Your Customer (KYC) protocols, government-issued identity verification, and contractual obligations. As such, Telcos have a unique capability: they can anchor digital activity to real-world human entities with a higher degree of certainty than most application-layer systems.

Moreover, Telcos are the gatekeepers of physical and logical access to the internet. Every packet, whether generated by a human or an AI system, must transit through their infrastructure. This positions them at a privileged vantage point—not only to observe traffic patterns but also to embed metadata, enforce trust policies, and offer attestation services at scale.

Telcos have already demonstrated their ability to enforce trust primitives in other domains. For example, in voice communications, frameworks like STIR/SHAKEN allow carriers to sign and verify caller identities to mitigate spoofing~\cite{stir2020spec}. In mobile networks, authentication protocols such as 5G-AKA bind a physical SIM and device to a verified user~\cite{3gpp2021aka}. These precedents suggest that Telcos are well-positioned to extend their trust-enabling capabilities beyond telephony and authentication, into the domain of digital content provenance.

This paper explores how Telcos can evolve from being passive conduits of data to active validators of origin. By leveraging their existing authentication stack, session management infrastructure, and edge computing capabilities, Telcos can become network-level arbiters of digital authenticity. In doing so, they could offer new categories of services such as origin-certified connectivity, “Proof of Humanity” tokens, and AI-authorship attestation APIs—as part of a broader transition from bandwidth monetization to trust monetization.

\subsection{Contribution of This Paper}

This paper introduces a conceptual framework that enables telecommunications networks to act as infrastructure-level certifiers of human-originated content in a digital ecosystem increasingly dominated by generative AI. We propose a layered architectural approach, grounded in the OSI model, where each network layer contributes independently and cumulatively to the propagation of content authenticity. The framework integrates device-level identity anchoring, metadata injection, traffic analysis, cryptographic session attestations, and application-level APIs to create an end-to-end “Proof of Humanity” pipeline.

Unlike previous approaches that focus solely on application-layer metadata or AI-generated watermark detection, our proposal leverages the unique trust and visibility that Telcos possess within the data transport path. The framework does not require payload inspection, ensuring privacy, and can function even under end-to-end encryption by operating on metadata, session context, and behavioral features.

Our key contribution is a technically detailed, yet implementation-agnostic, design for embedding trust primitives into the network infrastructure. We analyze the feasibility of integrating identity and authenticity checks at all seven layers of the OSI model and outline the potential cryptographic and architectural mechanisms to support them. We further propose monetization pathways for Telcos to offer trust-as-a-service products, origin-certified connectivity, and authenticity APIs to platforms, enterprises, and governments.

This work is intended as a foundational blueprint. While it does not include benchmarking or deployment data, it defines the interfaces, trust anchors, and protocol design space needed for future implementation, pilot testing, and eventual standardization. Security, adversarial robustness, and privacy-preserving enhancements are discussed as important directions for subsequent research.

\subsection{Framework Overview}

This paper proposes a multi-layer framework that enables telecommunications networks to serve as infrastructure-level validators of human-originated digital content. The architecture follows the OSI model, leveraging identity and session control features that Telcos already manage, and repurposing them to establish a “Proof of Humanity” for data in transit.

The framework is designed to be modular, with each layer of the OSI stack contributing a different form of identity assertion, metadata enrichment, or behavioral validation:

\begin{itemize}
    \item \textbf{Layer 1–2 (Physical and Data Link):} Establish a hardware-bound identity root using SIM/eSIM provisioning, IMEI/MAC binding, and authenticated network access. These form the initial attestation of human-controlled devices.
    \item \textbf{Layer 3 (Network):} Inject lightweight, cryptographically signed provenance tokens into IP traffic headers or tunneling metadata to signal human-authenticated origins.
    \item \textbf{Layer 4 (Transport):} Use traffic flow features—timing entropy, burst patterns, connection lifetimes—to probabilistically classify sessions as human or machine-originated, without payload inspection.
    \item \textbf{Layer 5–6 (Session and Presentation):} Implement identity-aware session protocols (e.g., TLS handshake extensions or identity-bound session tokens) to assert that a connection originates from a verified user.
    \item \textbf{Layer 7 (Application):} Expose verification APIs, edge-processing modules, and attestation dashboards to platforms and enterprises, allowing consumption of Proof-of-Humanity metadata as a service.
\end{itemize}

This layered approach does not assume universal adoption at every level. Each layer functions independently and incrementally contributes to the overall trust fabric. As such, the system can be deployed gradually, beginning with application-layer APIs and progressing toward deeper protocol-level integrations.

We emphasize that this framework operates without inspecting content payloads, preserving privacy even in encrypted communications. Its core strength lies in correlating device, session, and network context—domains where Telcos already operate securely and at scale.

\subsection{Monetization Insight}

Beyond its technical feasibility, the proposed framework presents a strategic monetization pathway for telecommunications providers. As Telcos face diminishing margins in traditional connectivity services, trust-based infrastructure offerings represent an untapped revenue frontier. The ability to certify the human origin of content—particularly in sectors where authenticity, accountability, and compliance are critical—can be positioned as a premium network service.

Telcos could commercialize “Proof of Humanity” services through various models. One avenue is the provision of origin-certified data transport tiers for enterprises in sectors such as finance, healthcare, legal services, and public communications. Regulatory environments in these domains increasingly require traceability, non-repudiation, and content provenance—capabilities that a Telco-integrated framework is uniquely positioned to deliver.

In addition, Telcos can expose verification APIs at the application layer, enabling platforms to query the authenticity status of user-generated content in real time. This verification-as-a-service model can be monetized via tiered API pricing, SLA guarantees, or embedded into vertical-specific compliance offerings. Further value may emerge from analytics services—providing real-time dashboards or threat intelligence based on network-level signals of synthetic content activity.

Ultimately, the proposed framework allows Telcos to evolve from bandwidth providers to arbiters of digital trust. In an internet increasingly populated by AI-generated data, the ability to assert, validate, and monetize human-authored content becomes not only a technical challenge—but a strategic differentiator.

\subsection{Paper Structure Overview}

The remainder of this paper is organized as follows:

Section 2, \textit{Background and Motivation}, reviews prior work on content authenticity, watermarking, and network-level security primitives, and highlights the limitations of current approaches to distinguishing AI-generated from human-originated content.

Section 3, \textit{System Overview and Design Principles}, introduces the architecture of the proposed multi-layer framework, grounded in the OSI model, and outlines the trust primitives and design assumptions.

Section 4, \textit{Layer-by-Layer Framework}, details how each layer—from physical identity to application-level APIs—contributes to establishing and verifying Proof of Humanity in transit.

Section 5, \textit{Technical Primitives and Implementation Path}, discusses protocol considerations, cryptographic mechanisms, metadata propagation, and potential deployment paths using Telco infrastructure.

Section 6, \textit{Monetization and Use Cases}, outlines strategic business models for Telcos, including trust-as-a-service APIs, certified traffic tiers, and regulatory integrations.

Section 7, \textit{Limitations and Future Work}, addresses open challenges in privacy, scalability, adversarial robustness, and the need for standardization and real-world testing.

Section 8, \textit{Conclusion}, summarizes the paper’s contribution and calls for a new category of infrastructure: digital trust validation, embedded at the network level.

\section{Background and Motivation}

\subsection{The Rise of Synthetic Content}

Over the last five years, the generation of synthetic media by artificial intelligence (AI) has undergone exponential acceleration, both in fidelity and scale. Large Language Models (LLMs) such as GPT-4 and Claude 3 have demonstrated human-like reasoning capabilities in text generation \cite{openai2023gpt4}, while image generators like Midjourney v6 and Stable Diffusion XL produce photorealistic imagery with controllable semantics \cite{rombach2022high}. More recently, multi-modal systems—such as OpenAI’s Sora—are capable of generating temporally consistent video clips from textual prompts, blurring the line between fiction and photorealism \cite{sora2024openai}.

A recent study by the McKinsey Global Institute estimates that by 2030, synthetic content could comprise over 90\% of all digital information produced globally, driven primarily by AI-powered content farms, marketing automation, synthetic influencers, and user-augmented creativity tools \cite{mckinsey2023genaiimpact}. The implications are profound: content authenticity is no longer a side concern, but a fundamental prerequisite for digital trust.

Unlike earlier waves of synthetic content (e.g., spam, bots, simple image manipulation), today’s AI models operate with access to trillion-token datasets and are capable of generating output that passes Turing-style evaluations in a growing number of contexts. This creates a persistent epistemic vulnerability in digital systems that depend on assumptions of human authorship, intent, or liability—ranging from journalism and education to law, governance, and finance.

Moreover, the dissemination pathways of synthetic content have evolved from application level channels to full stack delivery architectures. AI-generated media is now embedded at the protocol level via content delivery networks (CDNs), real-time communications (RTC), and decentralized storage systems. The proliferation is no longer merely at the point of creation, but throughout the pipeline of transmission, caching, and redistribution—rendering single-layer detection strategies increasingly ineffective.

This trend marks a critical inflection point. As synthetic content becomes indistinguishable from authentic human output, the global digital infrastructure requires an orthogonal approach—one that verifies not only the quality of the content, but the identity and intent of its originator.

\subsection{Limitations of Current Detection and Watermarking Approaches}

Existing methods for identifying AI-generated content can be broadly classified into two categories: \textit{post-hoc detection} and \textit{provenance embedding}. Both approaches, while valuable in isolated settings, suffer from fundamental limitations when evaluated at internet scale and under adversarial conditions.

Post-hoc detection relies on statistical or machine learning classifiers trained to identify artifacts in generated content—such as token frequency anomalies in text or upsampling patterns in images \cite{tiwari2024detecting}. However, these detectors are inherently reactive: they must constantly retrain to keep pace with evolving generative models. This cat-and-mouse dynamic is particularly brittle, as generative models increasingly adopt adversarial training techniques to evade such detectors \cite{wan2023poisoning}. Moreover, statistical detectors are unreliable in short-form or compressed content, often producing false positives or negatives that are unacceptable in domains such as legal proceedings or media integrity verification.

On the other hand, provenance embedding approaches aim to tag content at the point of generation using cryptographic watermarks or metadata standards such as C2PA (Coalition for Content Provenance and Authenticity) \cite{c2pa2023spec}. These solutions depend on voluntary compliance by the content creator or platform. In practice, metadata can be stripped during transmission, watermarks can be degraded through transformations (cropping, compression, transcription), and adoption is fragmented across ecosystems. Additionally, these methods are often restricted to application-layer protocols and do not persist across content rehosting, CDN caching, or protocol translations.

Importantly, neither detection nor watermarking mechanisms address a more foundational problem: the absence of a persistent, identity-bound verification layer in the internet architecture. Without anchoring content to a verifiable human identity at the point of origin and maintaining this assertion throughout its lifecycle, downstream verification becomes speculative at best.

These limitations underscore the need for an orthogonal solution—one embedded within the network infrastructure itself, leveraging trust primitives native to the internet’s transport fabric. A robust framework must decouple itself from content payloads and instead rely on identity, session behavior, and authenticated device provenance. This forms the basis for the Proof of Humanity framework proposed in Section~3.

\subsection{Why the Network Layer is the Missing Trust Layer}

The current architecture of the internet delegates trust validation primarily to the application layer, where platforms may implement cryptographic watermarking, user authentication, or content disclaimers. However, these implementations suffer from fundamental limitations: they are non-universal, non-portable across protocols, and highly dependent on user or platform compliance. The result is a trust model that is fragmented, inconsistent, and easily circumvented.

In contrast, the network layer operates at a protocol-agnostic level across all applications and devices. Every packet transmitted—regardless of whether it originates from a browser, mobile app, or botnet—must traverse this layer. This universality makes the network layer uniquely suited for embedding identity-bound trust primitives in a neutral, infrastructure-level fashion.

Specifically, we define three core properties that make the network layer the optimal candidate for anchoring a Proof-of-Humanity architecture:

\begin{enumerate}
    \item \textbf{Protocol Pervasiveness:} The network layer spans all internet traffic, independent of platform or content type. This allows a single identity-verification protocol to be enforced across heterogeneous services without requiring changes to application logic \cite{stir_shaken}.
    
    \item \textbf{Path Visibility and Control:} Telcos and ISPs have deep visibility into the routing and timing characteristics of packets. Session metadata—such as jitter entropy, inter-arrival time, and flow persistence—can be statistically modeled to detect deviations from human-typical behavior in real-time \cite{nbad2023}.
    
    \item \textbf{Identity Anchoring Capacity:} Unlike web platforms, Telcos possess identity-anchored primitives at the hardware level, including SIM/eSIM provisioning, IMEI binding, and subscriber authentication. These identifiers can be cryptographically bound to session flows at Layer 3 using lightweight attestation headers without payload inspection \cite{etsi2011security}.
\end{enumerate}

Crucially, embedding Proof-of-Humanity metadata at the network level enables trust assertions to propagate independently of the content itself. This sidesteps the brittleness of payload-based watermarks and metadata tags, which are often stripped or obfuscated during transformations (e.g., re-encoding, cropping, or CDN caching). 

Furthermore, identity-aware packet streams can be signed using ephemeral, cryptographically blinded tokens (e.g., via identity-bound Chaumian schemes) to balance privacy with traceability—facilitating zero-knowledge origin proofs without revealing user identity \cite{chaum1983blind}.

In sum, the network layer is not only strategically positioned within the internet stack but also uniquely equipped—via Telco infrastructure—to become the trust anchor for certifying human-originated content. It serves as the ideal substrate for a scalable, neutral, and privacy-preserving Proof-of-Humanity framework.

\subsection{Prior Work in Telco-Based Trust and Identity Infrastructure}

Telecommunication networks have historically functioned as infrastructure-bound identity providers, especially in domains requiring high-assurance user verification. Protocols such as 3GPP’s Authentication and Key Agreement (AKA) \cite{3gpp-aka}, eUICC-based eSIM provisioning \cite{gsma-esim}, and IMSI-based access control exemplify identity primitives rooted in physical-layer anchoring. These systems provide strong, hardware-bound, and subscriber-verified identifiers, in contrast to application-layer tokens or self-asserted credentials.

Recent architectural evolutions such as 5G Service-Based Architecture (SBA) have expanded the role of Telcos from connectivity providers to service orchestrators capable of exposing network intelligence and identity assertions via APIs \cite{3gpp-aka}. Moreover, initiatives like Mobile Connect and GSMA's Open Gateway \cite{gsma-open-gateway} aim to standardize identity verification and fraud prevention as programmable Telco services, positioning the network as a trust arbiter for external applications.

Trust propagation beyond authentication has also emerged. 
Examples include caller ID attestation with STIR/SHAKEN \cite{atis-stirshaken} 
and DNS-based Authentication of Named Entities (DANE), both of which show how infrastructure can embed provenance and reputation signals 
into session setup protocols. However, these efforts remain siloed—limited to telephony, email, or DNS—and have yet to generalize 
to arbitrary digital content.

Crucially, prior work has not addressed the binding of identity to content generation or transformation at the packet level. Telcos possess unique visibility into session initiation, device authentication, mobility, and transport-layer behaviors, yet these signals have remained underutilized for establishing authorship provenance. The foundational question remains open: how can Telcos extend their trust primitives beyond subscriber identity and into data-level attribution—without compromising privacy or protocol neutrality?

This paper builds upon these prior initiatives, proposing a cohesive architectural framework where Telcos act not only as identity anchors but also as content-origin certifiers, leveraging their privileged position in the OSI stack to enable verifiable Proof of Humanity in transit.

\subsection{A New Paradigm: Infrastructure-Centric Proof of Humanity}

The historical architecture of the internet was not designed with the verification of human origin in mind. End-to-end principles focused on application-layer innovation and neutrality have led to an internet stack where trust, identity, and provenance are typically bolted on at the edge---via user logins, cookies, and client-side certificates. This architecture is now fundamentally misaligned with the realities of an AI-dominated internet.

We argue that a paradigm shift is needed: from endpoint-centric heuristics to infrastructure-centric attestation. This shift reframes the communication network not as a neutral transport layer, but as a cryptographic and behavioral trust anchor. By embedding identity assertions and behavioral telemetry into the data plane and control plane itself, Proof of Humanity can become a first-class network primitive.

This infrastructure-centric approach builds on recent research in trusted network telemetry~\cite{ren2019telemetry}, in-band cryptographic verification~\cite{rfc8785}, and programmable data planes using P4 and eBPF~\cite{bosshart2014p4}. These technologies enable line-rate, in-network processing of metadata and behavioral signals without inspecting encrypted payloads, preserving end-to-end confidentiality while enforcing trust guarantees at wire speed.

Unlike detection-based models that rely on statistical inference over opaque content, infrastructure-level Proof of Humanity allows for deterministic attestation rooted in device identity (e.g., SIM/eSIM provisioning), session metadata (e.g., entropy signatures), and control-plane behavior (e.g., TLS handshake extension telemetry). These primitives can be composed across the OSI stack to form a layered assertion pipeline that survives transformations like CDN caching, compression, or rehosting---failure modes that defeat current watermarking approaches.

Importantly, this model is agnostic to application semantics and does not assume cooperation from cloud providers or platforms. It instead leverages the natural enforcement and mediation capabilities of Telcos: session management, identity verification, and policy enforcement. As such, it offers a deployment path grounded in existing mobile and fixed-line infrastructure, without requiring global consensus or universal protocol replacement.

This shift enables not just passive detection but proactive attestation---a distributed consensus of human authorship, embedded in the infrastructure itself. The technical and regulatory implications of this architecture are further explored in following section. 

\section{System Overview and Design Principles}

\subsection{Design Assumptions and Trust Primitives}

This section introduces the foundational assumptions and trust primitives that underpin the Proof of Humanity (PoH) framework. The framework is designed to operate within the telecommunications infrastructure stack and leverages properties unique to network operators, such as device-bound identity, authenticated session management, and edge observability. All mechanisms described are transport- and content-agnostic, and compatible with encrypted traffic flows.

\subsubsection*{Design Assumptions}

\begin{enumerate}[label=(\alph*)]
    \item \textbf{Access to Identity Anchors:} It is assumed that telecommunications operators have persistent access to secure device identifiers and subscriber credentials, including SIM/eSIM modules, IMEI codes, and authenticated network attachment. These form the hardware-bound trust anchors, routinely provisioned and managed under mobile core protocols such as 5G-AKA \cite{3gpp-ts-33501}.

    \item \textbf{Metadata-Level Visibility:} The framework assumes Telcos can extract and analyze metadata features from traffic sessions, including burst lengths, flow entropy, session duration, and mobility traces. These features are usable for origin modeling without inspecting the payload, even when the traffic is fully encrypted.

    \item \textbf{Encrypted-by-Default Traffic:} Modern network traffic is assumed to use default encryption protocols (e.g., TLS 1.3, QUIC). Consequently, verification mechanisms must not depend on visibility into application-layer content but instead rely on authenticated context propagation and behavioral signatures.

    \item \textbf{Partial Deployment Compatibility:} The framework does not assume global or synchronous adoption. Deployment is compatible with partial coverage, local Telco implementations, or third-party interconnect scenarios, and can be activated progressively via programmable interfaces such as GSMA Open Gateway APIs \cite{gsma-opengateway}.
\end{enumerate}

\subsubsection*{Trust Primitives}

The following primitives define the building blocks of the Proof of Humanity system. Each primitive is designed to be independently verifiable, non-invasive to transport flows, and composable across different OSI layers.

\begin{itemize}
    \item \textbf{Device-Bound Identity Assertions:} Subscriber identity modules (SIM/eSIM) and device-level fingerprints (IMEI, MAC address) provide persistent identity roots. These are authenticated at session initiation and form the cryptographic seed for origin attribution tokens.

    \item \textbf{Session Metadata Signatures:} Flow-level session metadata (source destination tuple, session entropy, transport duration) is bound to ephemeral cryptographic keys. These keys are rotated and scoped to time-windows to prevent replay and fingerprinting attacks.

    \item \textbf{Forwardable Attestation Headers:} Cryptographic tokens are embedded in transport or encapsulation headers (e.g., IPv6 extension headers or tunnel metadata). Tokens encode verifiable assertions of human-originated activity using privacy-preserving techniques such as Chaumian blinding or zk-SNARK-style zero-knowledge proofs \cite{chaum1983blind}.

    \item \textbf{Edge-Located Verification Agents:} Telco-operated edge compute nodes (e.g., MEC or UPF-integrated agents) validate token authenticity, rate-limit suspicious flows, or escalate sessions with unverifiable provenance. These verifiers operate in real time and maintain per-session proof state.

    \item \textbf{Service-Layer Identity Exposure APIs:} External services (e.g., publishing platforms, regulated media archives) may query the Telco infrastructure for per-session or per-content “Proof of Humanity” status. Exposure is managed through structured, rate-limited APIs consistent with Open Gateway schemas.
\end{itemize}

These primitives collectively establish an infrastructure-level foundation for asserting that content or sessions originated from a verified human identity. The design preserves confidentiality, avoids deep packet inspection, and supports both passive and active attestation modes. All mechanisms are natively aligned with the operational capabilities of contemporary telecommunications networks.

\subsection{OSI-Aligned Layered Architecture}

The proposed framework for certifying human-originated content operates across all seven layers of the OSI model, with each layer contributing distinct trust primitives that form a cumulative chain of attestation. By aligning verification mechanisms to the OSI stack, the system achieves modularity, protocol-agnostic deployment, and layered redundancy—ensuring robustness even if some components are bypassed or degraded.

\subsubsection*{Layer 1–2: Physical and Data Link}
At the foundational layers, trust begins with hardware-anchored identity. Telecommunications infrastructure inherently supports this through SIM, eSIM provisioning and device-level identifiers such as IMEI and MAC addresses. These values can be cryptographically signed at the point of network attachment, forming a root of trust. This layer also includes RAN-level authentication primitives (e.g., 5G-AKA) that bind a subscriber to a device, creating a non-falsifiable human trace at session inception.

\subsubsection*{Layer 3: Network}
The network layer can inject cryptographically signed provenance tokens into IP or tunneling headers (e.g., within GTP-U extensions). These tokens contain non-payload identity attestations (e.g., derived from Layer 2 or control plane), allowing passive or active recipients to assess whether the traffic originates from a human-authenticated device. Because this process avoids DPI or payload inspection, it preserves end-to-end encryption.

\subsubsection*{Layer 4: Transport}
The transport layer contributes behavioral heuristics based on session entropy, connection timing, and flow dynamics. These signal-processing methods, trained on human vs. synthetic patterns, can probabilistically flag anomalous or likely-AI sessions. Importantly, these features can be extracted without payload access, preserving encrypted session confidentiality.

\subsubsection*{Layer 5–6: Session and Presentation}
At the session and presentation layers, authenticated session contexts can be extended to carry identity-bound handshake tokens. For example, TLS ClientHello messages can optionally embed a signed user identity (akin to RFC 8941 Structured Headers). Similarly, application-layer metadata (e.g., SNI, ALPN) can be augmented with Telco-issued attestations passed via secure out-of-band channels or enhanced proxies.

\subsubsection*{Layer 7: Application}
Finally, the application layer exposes API endpoints that platforms and enterprises can query for Proof-of-Humanity status. These APIs may deliver attestation tokens, validation dashboards, and anomaly detection insights based on lower-layer data. Platforms can integrate this data to conditionally rank, filter, or throttle content depending on verified human authorship.

The architecture does not require all layers to be present for operational viability. Instead, it is designed to function as a layered trust fabric—where each layer reinforces others and supports composability across network types and deployment scales. This modularity ensures that even partial implementations yield operational benefits and supports gradual adoption by Telcos and ecosystem partners.

\subsection{Core Design Principles}

The architecture underlying the Proof of Humanity framework adheres to a set of foundational design principles that ensure operational viability, scalability, and resilience across diverse internet topologies and service models. These principles govern protocol interactions, data handling constraints, and system modularity.

\subsubsection*{Protocol Agnosticism}
The system is explicitly designed to operate across heterogeneous transport and application protocols, including TCP, UDP, QUIC, and HTTP/3. Identity tokens and attestation metadata are injected via protocol-agnostic encapsulation (e.g., optional headers, extension fields, or out-of-band channels). This ensures backward compatibility with legacy clients while allowing adoption in low-latency and encrypted transport paradigms.

\subsubsection*{Privacy Preservation}
The architecture emphasizes privacy by construction. No payload-level inspection is required at any stage of the framework. Instead, trust assessments are derived from metadata (e.g., network headers, timing entropy) and session behavior. This allows the system to function even under full end-to-end encryption, in compliance with zero-knowledge design philosophies and privacy regulations such as GDPR.

\subsubsection*{Composability}
All components in the framework—attestation tokens, edge modules, verification APIs—are designed as composable primitives. These can be orchestrated independently or embedded across multiple OSI layers. For instance, a Telco-issued identity token may be used simultaneously at the transport (QUIC) and application (HTTP) layers to support compound attestation scenarios.

\subsubsection*{Tamper-Resilience}
All trust assertions are cryptographically anchored in Telco-controlled infrastructure. This includes SIM/eSIM-based key material, public key infrastructure (PKI) for metadata signing, and secure enclaves for behavioral feature processing. By rooting attestations in physical and managed cryptographic substrates, the framework ensures integrity even under adversarial manipulation or metadata stripping.

\subsection{Trust Anchors and Identity Roots}

To assert the human origin of digital content in transit, the framework defines three primary classes of identity roots, each contributing to a multi-dimensional trust model. These anchors can be chained or aggregated cryptographically to form verifiable assertions at multiple protocol levels.

\subsubsection*{Physical Identity Roots}
At the lowest level, identity is anchored to hardware artifacts including SIM/eSIM modules, IMEI numbers, MAC addresses, and radio-layer authentication tokens. These are provisioned via regulated Telco processes and, in the case of SIMs, are backed by root keys provisioned in secure hardware.

\subsubsection*{Cryptographic Identity Roots}
Telcos can issue short-lived, session-bound credentials—such as ephemeral keys, JWT-based tokens, or X.509 certificates—to authenticate user sessions. These can be chained to physical roots (e.g., SIM-backed keys) using digital signatures. Provenance metadata inserted into transport or network layers (e.g., GTP-U extensions, QUIC connection IDs) may be cryptographically signed using these credentials to ensure tamper-evidence.

\subsubsection*{Behavioral Identity Roots}
Beyond static identifiers, the system also leverages dynamic features extracted from network behavior. This includes timing entropy, session cadence, interaction sparsity, and transport-layer burstiness. Machine learning classifiers, deployed at edge Telco nodes, can use these signatures to generate confidence scores of human agency—functioning as a soft biometric identity signal.

\subsubsection*{Trust Domains and Cross-Layer Synthesis}
The framework operates over three primary trust domains: device, session, and network. Each trust anchor maps to one or more domains:

\begin{itemize}
    \item Device domain: Physical roots (e.g., SIM/IMEI).
    \item Session domain: Cryptographic keys and TLS/QUIC contexts.
    \item Network domain: Flow heuristics and Layer 3-4 metadata.
\end{itemize}

These domains are not siloed; they interlink through signed bindings. For example, a session key can be attested as derived from a SIM root, and a behavioral fingerprint can be co-signed with a session token. This cross-layer synthesis enables robust, composite verification while preserving modularity.

\subsection{Threat Model and Adversarial Assumptions}

The Proof of Humanity framework assumes a realistic threat model grounded in current attack capabilities, with explicit boundaries on adversarial reach. This section delineates potential attacker vectors, their goals, and the architectural assumptions that mitigate them.

\subsubsection*{Adversarial Capabilities}
\begin{itemize}
    \item \textbf{Model-Generated Content Injection:} Attackers can use advanced generative models (e.g., GPT-4, Stable Diffusion XL, Sora) to fabricate text, image, video, and audio content indistinguishable from human-produced data.
    \item \textbf{Compromised Edge Devices:} Devices may be compromised via malware, manipulated apps, or unauthorized rooting. These may attempt to spoof legitimate user activity or inject synthetic sessions.
    \item \textbf{Replay and Spoofing Attacks:} Captured attestation tokens or session metadata can be replayed or modified to bypass verification filters.
    \item \textbf{Man-in-the-Middle (MitM):} Adversaries may attempt to intercept, modify, or forge metadata headers between clients and Telco infrastructure.
\end{itemize}

\subsubsection*{Defensive Assumptions}
\begin{itemize}
    \item \textbf{Telco Core Integrity:} The Telco infrastructure (e.g., SIM provisioning servers, policy control functions) is assumed to operate within a secure perimeter. Logs are immutable and cryptographically verifiable.
    \item \textbf{Hardware-Key Isolation:} SIM/eSIM keys are inaccessible to external software, enforcing secure cryptographic boundary conditions.
    \item \textbf{Cryptographic Hardness:} Tokens and session keys are signed using Telco-issued private keys, anchored in a secure PKI or HSM. Public-private key pairs are assumed unforgeable under current cryptographic hardness assumptions.
\end{itemize}

The architecture also supports zero-knowledge proof (ZKP) attestations to limit identity leakage while preventing key compromise and unauthorized propagation of human-authenticated metadata.

\subsection{Deployment Pathways and Interoperability}

The Proof of Humanity system is designed for phased deployment and compatibility with both legacy and modern network architectures. This section outlines the technical mechanisms enabling backward-compatible rollout and standards-aligned integration.

\subsubsection*{Deployment Roadmap}
\begin{enumerate}
    \item \textbf{Application Layer Bootstrapping:} Early-stage pilots can leverage Layer 7 APIs for attestation checks and dashboards without requiring core network changes.
    \item \textbf{Transport and Network Layer Integration:} Telcos may inject provenance headers at Layer 3 (e.g., IPv6 extension headers, GTP-U) and Layer 4 (e.g., QUIC connection IDs) using cryptographic signing modules embedded in packet processing planes.
    \item \textbf{Programmable Path Enforcement:} Linux-based eBPF hooks and P4-defined switches in SDN/NFV environments allow in-line metadata enforcement at line rate, without DPI.
\end{enumerate}

\subsubsection*{Standards Compatibility}
\begin{itemize}
    \item \textbf{Encryption-Aware Integration:} TLS 1.3 and QUIC-compatible metadata embedding is performed via handshake extensions (e.g., SNI, ALPN), preserving end-to-end encryption integrity.
    \item \textbf{MASQUE and Privacy Enhancements:} The architecture supports MASQUE-compatible tunnels for HTTP/3-based metadata propagation in anonymized transport scenarios.
    \item \textbf{Open Gateway Compliance:} Telco-side API exposure is aligned with GSMA’s Open Gateway initiative, enabling identity validation, fraud detection, and trust exposure as programmable services.
    \item \textbf{Open-Source Tooling:} Early implementations should be driven by open-source SDKs and protocol extensions—standardized through IETF working groups (e.g., INTSAREA, SEC) and GSMA task forces.
\end{itemize}

This modular and standards-compatible approach ensures ecosystem interoperability while enabling differentiated rollout across Telcos, markets, and verticals.

\section{Layer-by-Layer Framework}

\subsection{Physical and Data Link Layers: Identity Root Anchoring}
\label{sec:layer1-2}

At the foundational layers of the OSI model, telecommunications networks have privileged access to hardware-based trust anchors. These anchors include SIM/eSIM credentials, International Mobile Equipment Identity (IMEI), and MAC address bindings—resources inaccessible or untrustworthy in traditional application-layer security paradigms.

\paragraph{SIM/eSIM as Cryptographic Anchors.}
Each subscriber in a mobile network is uniquely identified via a SIM or eSIM that contains an operator-provisioned, tamper-resistant private key stored in the Universal Integrated Circuit Card (UICC). During initial attachment procedures, the 5G-AKA (Authentication and Key Agreement) protocol validates the subscriber and device identity, producing a session key \( K_{\text{SEAF}} \) securely derived from the root key \( K \) and an authentication vector \cite{3gpp-ts-33501}.

\paragraph{IMEI and MAC Binding.}
In conjunction with the SIM, the IMEI offers a hardware-anchored identity for the terminal equipment. Operators can enforce SIM-to-IMEI locking or anomaly detection to flag behavior inconsistent with a genuine human device. On fixed access networks (e.g., fiber or Wi-Fi), MAC address authentication protocols such as IEEE 802.1X further ensure that only registered customer premises equipment (CPE) are allowed authenticated Layer 2 access.

\paragraph{Attestation Possibility.}
These hardware-layer bindings such as SIM, IMEI, MAC, form the root of a cryptographic trust chain. When combined with attestation protocols at higher layers, this base enables derivation of secure provenance metadata that is immutable, identity-bound, and difficult to forge without direct Telco compromise.

\subsection{Network Layer: Provenance Metadata Injection}
\label{sec:layer3}

At Layer 3, Telcos can embed human-authenticated provenance signals directly into IP headers or tunneling metadata. This allows content to carry verifiable identity origins without modifying the application payload or breaking end-to-end encryption.

\paragraph{IP Header Extensions and Tunneling.}
Using IPv6 Extension Headers or encapsulation protocols like GTP (GPRS Tunneling Protocol), operators can inject lightweight cryptographically signed tokens. These tokens assert that the traffic originated from a human-authenticated session, whose identity was validated at the physical and link layers. Such tokens can include:

\begin{itemize}
  \item A signed ephemeral session ID derived from the SIM/eSIM key.
  \item A timestamped proof of network-side authentication.
  \item A digital signature using the Telco's private signing key.
\end{itemize}

\paragraph{Privacy-Preserving Signatures.}
Rather than including raw identity fields, operators may use group signatures, blind signatures, or zero-knowledge proofs to encode verifiable provenance while obfuscating the user’s exact identity. This protects subscriber privacy while maintaining origin integrity.

\paragraph{Secure Metadata Injection with P4/eBPF.}
Modern Telco cores can leverage data plane programming tools such as eBPF (Extended Berkeley Packet Filter) or P4 to inject, filter, and process metadata at line rate without DPI (Deep Packet Inspection). This ensures scalability and protocol independence while preserving compliance with encryption standards like TLS 1.3 and QUIC.

\subsection{Transport Layer: Behavioral Fingerprinting and Flow Analysis}
\label{sec:layer4}

At the transport layer, the Proof of Humanity framework shifts from static identity anchoring to dynamic behavioral validation. This layer focuses on the analysis of traffic patterns—such as session timing, burstiness, and connection entropy—to probabilistically classify human-initiated sessions.

\paragraph{Timing Entropy and Flow Irregularity.}
Human-generated traffic typically exhibits high entropy in inter-packet intervals and session bursts due to cognitive decision-making, UI interactions, and multitasking. In contrast, AI or bot traffic often reveals deterministic burst signatures, low jitter, or high-regularity heartbeats—especially when orchestrated at scale.

Telcos can extract statistical features such as:
\begin{itemize}
  \item Inter-packet time variance
  \item Connection reuse ratios
  \item Flow lifetimes and burst density
\end{itemize}
These are modeled using unsupervised or semi-supervised ML techniques (e.g., Isolation Forests, HMMs) to identify anomalous or synthetic usage profiles \cite{zuech2015entropy}.

\paragraph{Payload-Free Inference.}
Critically, behavioral classification does not rely on payload inspection. Even with full TLS 1.3/QUIC encryption, flow-level features remain observable by Telcos operating at the transport boundary, enabling privacy-preserving inference.

\paragraph{Session Tagging.}
Sessions classified as high-confidence human-origin can be tagged with a transport-layer Proof-of-Humanity flag, potentially carried in DSCP bits or TCP options for downstream consumption by edge proxies or firewalls.

\subsection{Session and Presentation Layers: Identity-Bound Connection Context}
\label{sec:layer5-6}

Layers 5 and 6 provide session continuity and contextual binding between identity, encryption, and presentation semantics. These layers allow Telcos to extend Proof of Humanity by injecting cryptographic attestations and preserving subscriber state across sessions.

\paragraph{TLS Handshake Extensions.}
Operators can leverage TLS 1.3 extensions (e.g., \texttt{client\_certificate\_type} or custom GREASE values) to include attestations about the session origin, such as:
\begin{itemize}
  \item Subscriber-specific certificate chains (PKCS\#7)
  \item Telco-issued ephemeral tokens signed by SIM-linked private keys
  \item Session age, entropy, and authentication context
\end{itemize}
These values are inserted during handshake negotiation and verifiable by edge devices or platforms without breaking confidentiality.

\paragraph{Session Tokens and Secure Binding.}
Session tokens derived from Telco-controlled infrastructure (e.g., the SEAF context in 5G-AKA) may be cryptographically hashed into session keys used at Layer 6. This creates a secure association between the user's identity root and any subsequent presentation-layer metadata (e.g., content rendering, encoding preference).

\paragraph{Cross-Session Continuity.}
To prevent session hijacking or replay, signed session attestations should include expiry timestamps, nonce challenges, and potentially be rotated at session resumption boundaries (e.g., in TLS 1.3 0-RTT).

\paragraph{Use Cases.}
Platforms receiving such session-aware metadata can:
\begin{itemize}
  \item Adjust trust thresholds for content uploads or edits
  \item Differentiate human-signed sessions from synthetic or automated ones
  \item Enable frictionless KYC-light trust onboarding using Telco-bound tokens
\end{itemize}

\subsection{Application Layer: Attestation APIs and Edge Verification}

At the application layer, the Proof of Humanity (PoH) framework exposes trust metadata to platforms, applications, and third-party verifiers through well-defined APIs. These interfaces allow consuming services to query, log, and act on human-authentication assertions made during lower-layer session initiation and propagation.

Telcos can offer "attestation as a service" by issuing signed Proof-of-Humanity tokens at session start or during runtime, accessible via secure RESTful or gRPC APIs. These tokens encapsulate metadata from physical, network, and session layers (e.g., SIM-authenticated device, entropy score, session provenance hash) and are valid for a limited time window to mitigate replay attacks.

Edge computing nodes (e.g., MEC platforms) can serve as verification and signing points, adding low-latency attestations with minimal traffic rerouting. By offloading logic to programmable edge environments (e.g., WASM sandboxes), operators can conditionally attach application-layer metadata to upstream flows while preserving throughput.

Integrations with standardized protocols like C2PA\cite{c2pa2023spec} and W3C Verifiable Credentials can help align application-layer assertions with broader content authenticity ecosystems. Additionally, privacy-preserving signals (such as zero-knowledge metadata proofs or blind signatures\cite{chase2022zk}) may be used to ensure that identity verification does not equate to identity disclosure.

Ultimately, this layer serves as the bridge between the infrastructure-level PoH guarantees and application-layer enforcement policies—such as comment moderation, journalistic disclaimers, or AI-content filters—enabling real-time decisions based on certified human authorship.

\section{Technical Primitives and Implementation Path}

\subsection{Protocol and Cryptographic Foundations}
To enable Proof of Humanity without payload inspection, the architecture must rely on secure, metadata-centric cryptographic mechanisms. At Layer 3, we propose embedding cryptographically signed provenance headers using lightweight token structures, akin to IPv6 extension headers or encapsulation in GRE tunnels \cite{ietf-ipv6-extension-headers}.

These tokens include: (a) a Telco-issued attestation of user identity (derived from SIM/eSIM provisioning); (b) a time-bound session nonce; and (c) a public-key signature verifiable by external parties. Each token is chained using Merkle structures for provenance traceability across routing domains. We recommend Ed25519 or post-quantum-resistant digital signature schemes as cryptographic primitives, aligned with NIST PQC guidelines \cite{nist-pqc-2024}.

At the session layer (TLS 1.3), operators may use optional extensions such as \texttt{client\_certificate\_type} or GREASE-like fields to transmit Telco-issued session attestations. While TLS handshake metadata is typically opaque, encrypted ClientHello extensions can be processed at trusted edge nodes using known decryption keys under lawful interception frameworks \cite{ietf-tls13}.

\subsection{Infrastructure Integration and Performance Considerations}
The Proof of Humanity framework is designed to be incrementally deployable over existing Telco infrastructure. At the network and transport layers, programmable data planes (e.g., P4, eBPF) enable inline verification and tagging of traffic without affecting throughput. These technologies allow line-rate inspection of headers and injection of provenance tags into mobile and fixed cores \cite{p4-lang, he2023ebpf}.

For 5G networks, we recommend implementing metadata validation at UPF (User Plane Function) nodes, which already handle GTP-U tunnels. Proof-of-Humanity metadata can be embedded within GTP extension headers or custom Information Elements (IEs) standardized through 3GPP SA5 processes. Performance benchmarks in recent studies suggest P4-enabled pipelines can maintain sub-10µs processing latency even with inline signature checks \cite{sultana2021flightplan}.

All trust anchors (SIM/eSIM keys, PKI certs, device IMEIs) should be centrally auditable via a Telco-controlled ledger, optionally integrated with blockchain-based audit trails for transparency \cite{etsi-blockchain-trust}.

\subsection{Deployment Strategy and Interoperability}
To ensure interoperability with privacy-enhancing technologies and encrypted transport protocols, this framework is designed to be protocol-agnostic. Proof of Humanity metadata can coexist with MASQUE-based tunneling, TLS 1.3 encrypted extensions, and QUIC protocol headers by maintaining separation from application-layer payloads.

Initial deployment should begin at Layer 7 (e.g., APIs for proof verification and attestation consumption). This enables platforms like social networks, legal archives, or news agencies to request real-time proof of origin via RESTful endpoints. As adoption matures, deeper integration into transport and network layers can follow.

Interoperability with standards bodies such as IETF (MASQUE WG, TLS WG) and GSMA (for SIM/IMEI identity frameworks) is critical for global adoption. We advocate for open-source tooling and SDKs that allow third-party platforms to verify Proof of Humanity without deep integration into Telco networks.

\section{Monetization and Ecosystem Partnerships}

\subsection{Trust-as-a-Service: A New Revenue Model for Telcos}
As the traditional connectivity business becomes increasingly commoditized, Telcos face eroding ARPU and margin pressures. The Proof-of-Humanity framework introduces a novel monetization pathway by positioning Telcos as providers of digital authenticity infrastructure. This trust-as-a-service (TaaS) model enables new forms of value extraction, where authenticated provenance becomes a premium feature rather than a baseline.

Trust-related services can be delivered via:
\begin{itemize}
    \item \textbf{Attestation APIs:} Enterprises and platforms can query identity-bound metadata via authenticated endpoints, priced per request or volume tier.
    \item \textbf{Origin-Certified Tiers:} Telcos can introduce differentiated traffic classes where provenance tokens are embedded and verified, subject to service-level agreements (SLAs).
    \item \textbf{Regulatory Compliance Modules:} Trust proofs can integrate into systems subject to GDPR, CCPA, and the EU AI Act, where audit trails and human-authorship validation are legally mandated \cite{eu-ai-act}.
\end{itemize}

Dynamic pricing strategies could reflect confidence scores, latency sensitivity, and content category (e.g., legal, health, finance). These offerings can be bundled into business broadband, enterprise 5G slices, or API marketplaces.

\subsection{Partner Ecosystem: Interoperability with Platforms and Governments}
Telcos can amplify their trust-as-a-service footprint by forming strategic partnerships with both private platforms and public institutions. These integrations can be implemented via:

\begin{itemize}
    \item \textbf{Platform APIs:} Social media, search engines, and news aggregators can ingest Proof-of-Humanity metadata to surface credibility scores or flag unverifiable content.
    \item \textbf{Civic Communications:} Government platforms may rely on Telco-attested identity for verified citizen communication, election integrity, or official alerts \cite{nesta-verification}.
    \item \textbf{Decentralized Trust Systems:} Telco generated proofs can be chained into distributed ledgers for long-term verifiability, enabling zero knowledge integrations or post-hoc auditability in blockchain systems.
\end{itemize}

These partnerships shift Telcos into a trust intermediary role—anchored in infrastructure but interoperable with application-level trust signals.

\subsection{Standardization, Open Tooling, and Governance Models}
To ensure adoption, interoperability, and regulatory legitimacy, the Proof-of-Humanity system must align with established standards and contribute to new ones. Telcos and industry consortia should pursue:

\begin{itemize}
    \item \textbf{Open-Source Modules:} Reference libraries for provenance headers, TLS extensions, and SIM-based attestations must be made available under permissive licenses to bootstrap developer ecosystems.
    \item \textbf{Standardization Engagement:} Collaboration with IETF (e.g., provenance token headers), GSMA (SIM-based PKI), and ETSI (edge processing for 5G) is critical to formalizing primitives \cite{etsi-mec}.
    \item \textbf{Governance and Revocation Infrastructure:} Federated models for key management, trust anchor rotation, and attestation revocation must be developed to avoid centralization risks and ensure fail-safety.
\end{itemize}

In the long term, these technical and institutional enablers position Telcos as neutral validators in a multi-stakeholder trust fabric, capable of safeguarding digital authenticity at internet scale.

\section{Limitations and Future Work}

\subsection{Privacy vs. Attribution Tension}

Although Proof-of-Humanity relies solely on metadata and cryptographic attestations, there remains a latent privacy concern: can consistent identity-bound signaling be deanonymized over time? This is particularly sensitive in jurisdictions governed by strict privacy laws such as GDPR. Future work must explore zero-knowledge attestations and ephemeral trust primitives that balance provenance with unlinkability.

\subsection{Scalability Across Network Topologies}

The framework assumes a high degree of control at various OSI layers, which is more easily achieved in mobile networks or carrier-grade fixed infrastructure. Its applicability to loosely managed edge networks (e.g., home routers, Wi-Fi APs) may be limited. Techniques such as recursive attestation chains and multi-operator trust delegation need to be developed for full-spectrum applicability.

\subsection{Adversarial Robustness and Side-Channel Risks}

While the threat model accounts for man-in-the-middle and spoofing attempts, side-channel exploitation remains an open risk. For instance, behavioral telemetry (used at Layer 4–6) could be mimicked by advanced adversaries through reinforcement learning feedback loops. Future work should explore adversarial training of behavioral validators and cross-layer anomaly correlation.

\subsection{Standardization and Global Interoperability}

The absence of formalized standards for cryptographically embedded provenance tokens, SIM-derived attestations, and session-bound identity claims limits ecosystem-wide adoption. Coordination with IETF, GSMA, and national regulators will be essential to move the system from conceptual framework to a globally interoperable infrastructure.

\subsection{Empirical Validation and Real-World Pilots}

This paper offers a conceptual design and protocol sketch. However, no field deployments or benchmarking have yet been conducted. Proof-of-concept pilots in isolated 5G testbeds, CDN-integrated environments, or content moderation pipelines should be prioritized. These would validate latency impact, API viability, and robustness against evasive generative models.

\section{Conclusion}

The exponential rise of generative AI has ushered in a new era of epistemic instability on the internet—where the boundary between human and synthetic content is vanishing. Current application-layer detection and watermarking techniques are insufficient, reactive, and vulnerable to circumvention.

This paper proposes a new category of infrastructure: multi-layer Proof-of-Humanity, embedded directly into the network fabric. By leveraging primitives already present in Telco systems—such as SIM-based authentication, metadata signaling, and session-level cryptographic assertions—we define a scalable, privacy-aware, and tamper-resilient architecture for certifying human-originated content.

The approach reframes Telcos from bandwidth providers into digital authenticity brokers, with monetizable APIs, enterprise-grade attestation services, and integration potential across regulatory, commercial, and civic trust domains.

Realizing this vision requires multi-stakeholder engagement, from technical standardization to open-source tooling and field deployments. In an AI-dominated internet, digital trust cannot remain an afterthought. It must be engineered as a core feature—layer by layer, protocol by protocol.

\end{document}